\documentclass[aps,prl,twocolumn,showpacs,floatfix]{revtex4}

\usepackage{graphicx} 
\usepackage{bm}

\begin{document}

\title{Interlayer Aharonov-Bohm interference in tilted magnetic fields
  in quasi-one-dimensional layered materials}

\author{Benjamin K. Cooper} \affiliation{Condensed Matter Theory
  Center and Center for Superconductivity Research, Department of
  Physics, University of Maryland, College Park, Maryland 20742-4111,
  USA}

\author{Victor M.~Yakovenko} \affiliation{Condensed Matter Theory
  Center and Center for Superconductivity Research, Department of
  Physics, University of Maryland, College Park, Maryland 20742-4111,
  USA}

\date{\bf cond-mat/0509039, v.2 13 December 2005}


\begin{abstract} 
  Different types of angular magnetoresistance oscillations in
  quasi-one-dimensional layered materials, such as organic conductors
  $\rm(TMTSF)_2X$, are explained in terms of Aharonov-Bohm
  interference in interlayer electron tunneling.  A two-parameter
  pattern of oscillations for generic orientations of a magnetic field
  is visualized and compared with the experimental data.  Connections
  with angular magnetoresistance oscillations in other layered
  materials are discussed.
\end{abstract} 

\maketitle

Angular magnetoresistance oscillations (AMRO), where resistivity
oscillates as a function of the magnetic field orientation, were
originally discovered \cite{Kartsovnik88,Kajita} in the
quasi-two-dimensional (Q2D) organic conductors of the (BEDT-TTF)$_2$X
family \cite{book}.  AMRO are distinct from the Shubnikov-de Haas and
de Haas-van Alphen oscillations and are now widely used for direct
mapping of the Fermi surfaces of layered materials
\cite{Kartsovnik92,Goddard}.  AMRO have been observed not only in many
organic conductors, but also in intercalated graphite \cite{Iye94},
$\rm Sr_2RuO_4$ \cite{Sr2RuO4-AMRO}, $\rm Tl_2Ba_2CuO_6$
\cite{Yakovenko99f,Hussey03}, and GaAs/AlGaAs superlattices
\cite{GaAs}.  Early theories of AMRO \cite{Yamaji,Yagi,Kartsovnik92}
were formulated in terms of semiclassical electron trajectories on a
cylindrical 3D Fermi surface.  Then it was realized that AMRO can
exist already for two layers \cite{Kurihara92,McKenzie,Osada03}, and
they represent an Aharonov-Bohm interference effect in interlayer
tunneling \cite{YC}.  Some experimental evidence for AMRO in
semiconducting bilayers has been found \cite{Boebinger}, but more
systematic measurements are necessary.

\begin{figure}
  \includegraphics[width=0.43\linewidth]{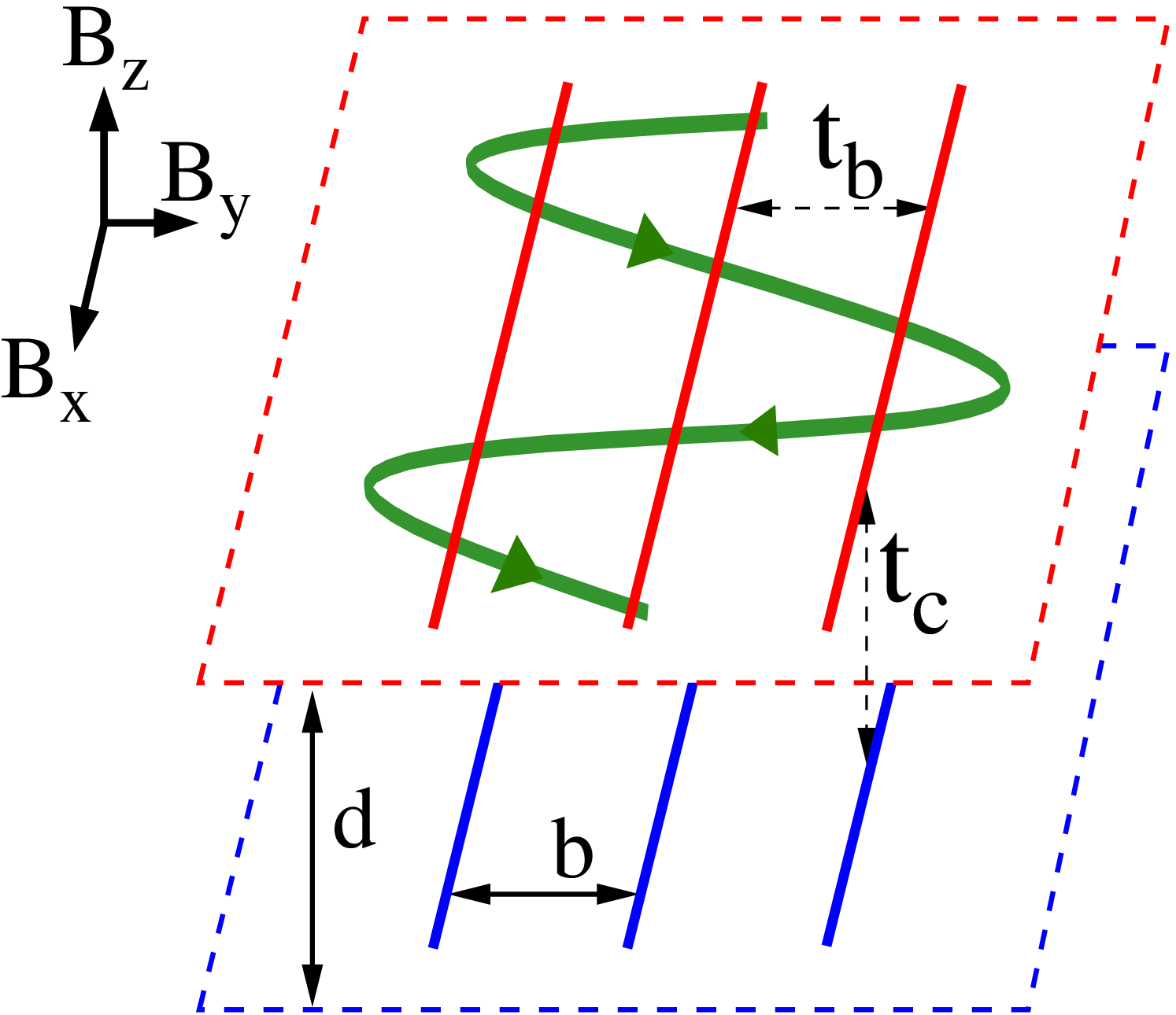} 
  \hfill
  \includegraphics[width=0.49\linewidth]{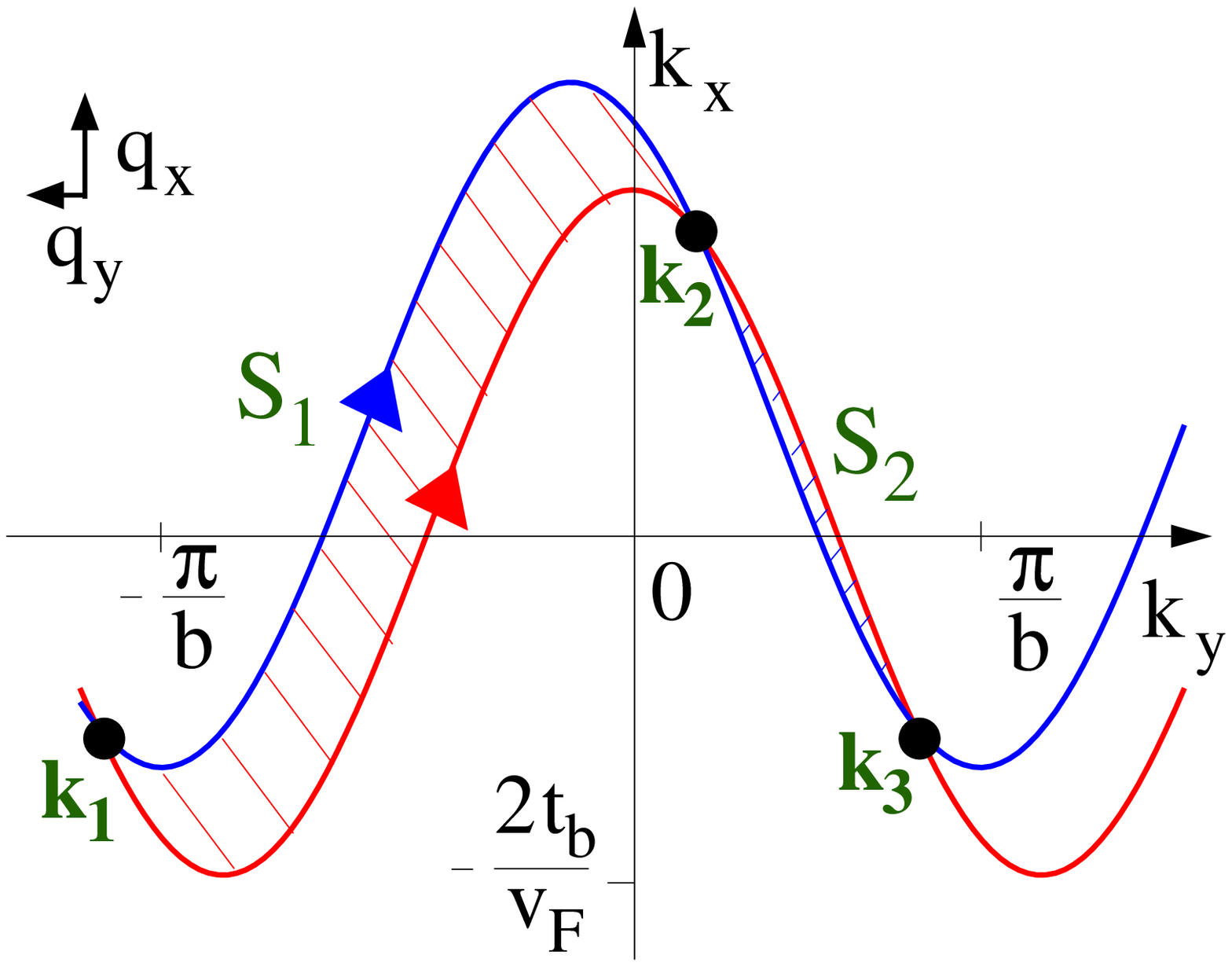}
\caption{(a) Geometry of electron tunneling 
  between two Q1D layers.  The sinusoidal line represents the in-plane
  electron trajectory; $t_b$ and $t_c$ are the amplitudes of
  interchain tunneling.  (b) The Fermi surfaces of the two layers,
  shifted by the vector ${\bm q}$ in Eq.\ (\ref{q}). The shaded areas
  $S_1$ and $S_2$ lead to interference oscillation in the presence of
  $B_z$.}
\label{fig:geometry}
\end{figure}

AMRO were also found in quasi-one-dimensional (Q1D) organic conductors
with open Fermi surfaces, such as $\rm(TMTSF)_2X$ \cite{book}.  These
materials consist of parallel chains along the $x$ axis, which form
layers with the interlayer spacing $d$ along the $z$ axis and the
interchain spacing $b$ along the $y$ axis, as shown in Fig.\ 
\ref{fig:geometry}a.  Originally, three different AMRO were discovered
in the Q1D conductors: the Lebed magic angles
\cite{Lebed86a,Osada91,Naughton91,Chaikin92c} for a magnetic field
rotation in the $(y,z)$ plane, the Danner, Kang, and Chaikin (DKC)
oscillations in the $(x,z)$ plane \cite{Chaikin94a,Chaikin95c}, and
the third angular effect in the $(x,y)$ plane
\cite{Yoshino95,Osada96,Lebed97b}.  Then Lee and Naughton
\cite{Naughton98a} found combinations of all three effects for generic
magnetic field rotations.  It became clear that all types of AMRO in
Q1D conductors have a common origin and should be explained by a
single unified theory.

In $\rm(TMTSF)_2X$, the in-plane tunneling amplitude between the
chains, $t_b\sim250$ K \cite{Chaikin94a,Chaikin95c}, is much greater
than the inter-plane tunneling amplitude $t_c\sim10$ K \cite{book}.
Thus, we can treat interlayer tunneling as a perturbation and study
the interlayer conductivity $\sigma_c$ between just two layers for a
tilted magnetic field ${\bm B}=(B_x,B_y,B_z)$, as shown in Fig.\ 
\ref{fig:geometry}a.  This bilayer approach \cite{McKenzie,Osada03}
only assumes phase memory of interlayer tunneling within a decoherence
time $\tau$ and does not require a well-defined momentum $k_z$ and a
coherent 3D Fermi surface.  It gives a simple and transparent
interpretation of the most general Lee-Naughton oscillations
\cite{Naughton98a} in terms of Aharonov-Bohm interference in
interlayer tunneling.  The results are equivalent to other approaches
based on the classical Boltzmann equation
\cite{Osada96,Lebed97b,Naughton98a,Osada99,Yoshino99} and the quantum
Kubo formula \cite{Osada92,Lebed03,Lebed05}.  We calculate a contour
plots of $\sigma_c$ as a function of two ratios $B_x/B_z$ and
$B_y/B_z$ for models with one or several interlayer tunneling
amplitudes \cite{Osada92,Chaikin02}.  This type of visualization
clearly reveals agreement and disagreement between theory and
experiment and allows to determine the electronic parameters of the
Q1D materials.  The results can be also applied to Q1D semiconducting
bilayers consisting of quantum wires induced by an array parallel
gates, as shown in Fig.\ \ref{fig:geometry}a.

Let us consider tunneling between two Q1D layers.  The in-plane
electron dispersion is
\begin{equation}
  \varepsilon(k_x,k_y)=\pm v_Fk_x-2t_b\cos(k_yb/\hbar), 
\label{E_k}  
\end{equation}
where energy $\varepsilon$ is measured from the Fermi energy, $\pm
v_F$ are the Fermi velocities on the opposite sheets of the open Fermi
surface, ${\bm k}=(k_x,k_y)$ is the in-plane momentum, and $k_x$ is
measured from the Fermi momentum.  The interlayer tunneling is
described by the Hamiltonian
\begin{eqnarray}
  && \hat{H}_{\perp} = t_c \int \hat\psi_2^\dag({\bm r})\,
  \hat\psi_1({\bm r})\, e^{i\phi({\bm r})} d^2r\,
  + \mbox{H.c.},
\label{H_perp} \\
  && \phi({\bm r})=\frac{ed}{\hbar c}A_z({\bm r}), \quad
  A_z({\bm r})=B_x y- B_y x,
\label{phi}
\end{eqnarray}
where ${\bm r}=(x,y)$, $c$ is the speed of light, $e$ is the electron
charge, $A_z$ is the vector potential, and $\hat\psi_{1,2}$ are the
electron destruction operators in the layers 1 and 2.  The gauge phase
$\phi({\bm r})$ is due to the in-plane magnetic field.

We treat the in-plane electron motion quasiclassically.  For
$B_z\neq0$, electrons move in time $t$ along sinusoidal trajectories
\cite{YG}, as shown in Fig.\ \ref{fig:geometry}a,
\begin{equation}
  x(t)=x_0 \pm v_Ft,\; y(t)=y_0\pm
  \left({2t_bc\over ev_FB_z}\right)\,\cos(\omega_ct).
\label{xy}
\end{equation}
Instead of the magnetic field components $(B_x,B_y,B_z)$, it is
convenient to introduce the variables $\omega_c$, $B_x'$ and $B_y'$
defined by the following relations:
\begin{equation}
  \omega_c={ebv_FB_z \over \hbar c}, \quad 
  B_x'={B_x\over B_z}{2t_bd\over \hbar v_F}, \quad 
  B_y'={B_y \over B_z}{d \over b}.
\label{omega_c}
\end{equation}
The cyclotron frequency $\omega_c$ is simply proportional to $B_z$,
whereas the dimensionless variables $B_x'$ and $B_y'$ are proportional
to the ratios of the magnetic field components
$B_x/B_z=\cos\varphi\tan\theta$ and $B_y/B_z=\sin\varphi\tan\theta$.
Although these ratios can be expressed in terms of the spherical
angles $\theta$ and $\varphi$, we believe that presentation and
visualization of the results using $B_x'$ and $B_y'$ is simpler and
more insightful than in the spherical angles
\cite{Osada96,Osada03,Osada99}.

The gauge phase (\ref{phi}) in Eq.\ (\ref{H_perp}) leads to
interference between interlayer tunneling amplitudes $t_c
e^{i\phi({\bm r})}$ along the trajectory ${\bm r}(t)$.  In Eq.\ 
(\ref{xy}), $y(t)$ oscillates with the period $\Delta
t=2\pi/\omega_c$, whereas $x(t)$ steadily increases, accumulating the
phase $\Delta\phi=edB_yv_F\Delta t/\hbar c$ over one period.  The
average $\langle e^{i\phi(t)}\rangle_t$ vanishes unless
$\Delta\phi=2\pi n$, where $n$ is an integer.  This condition selects
the Lebed magic angles $B_y'=n$ \cite{Lebed86a}, which in the
spherical coordinates are $\sin\varphi=n(b/d)\cot\theta$
\cite{Naughton98a}.  Using Eqs.\ (\ref{phi}), (\ref{xy}) and
(\ref{omega_c}), we find the effective interlayer tunneling amplitude
$\tilde t_c$
\begin{equation}
  \tilde t_c = t_c \left\langle e^{i\phi(t)}\right\rangle_t
  = t_c\,J_n(B_x') \quad \mbox{for}  \quad B_y'=n,
\label{J_n}
\end{equation}
where $J_n$ is the Bessel function.  

AMRO result from a periodic modulation of the effective interlayer
coupling $\tilde t_c$ in Eq.\ (\ref{J_n}) due to interlayer
Aharonov-Bohm interference.  The condition $B_y'=n$ requires that the
flux of $B_y$ through the area, formed by the interlayer distance $d$
and the electron trajectory period $\Delta x=v_F\Delta t$, is
$n\Phi_0$, where $\Phi_0=hc/e$ is the flux quantum.  In addition,
$\tilde t_c^2$ (\ref{J_n}) oscillates as a function of $B_x'$ with the
period $\Delta B_x'=\pi$.  These DKC oscillations \cite{Chaikin94a}
are related to the flux of $B_x$ through the area bounded by $d$ and
$\Delta y=4t_bc/ev_FB_z$, the transverse width of the electron
trajectory in Eq.\ (\ref{xy}).  More precisely, it is necessary to
consider the distance between the turning points of an electron
trajectory, as viewed along the vector $(B_x,B_y)$.  This will be
discussed in more detail from the momentum-space point of view.

\begin{figure}
  \includegraphics[width=0.8\linewidth]{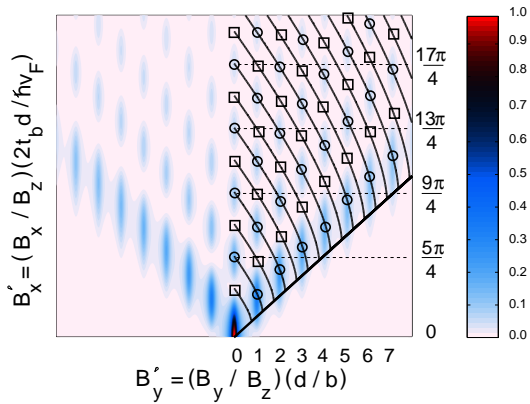}
\caption{The contour plot of angular oscillations in the 
  normalized interlayer dc conductivity $\sigma_c({\bm
    B},0)/\sigma_c(0,0)$, Eq.\ (\ref{sigma_c}).  The variables $B_x'$
  and $B_y'$ are defined in Eq.\ (\ref{omega_c}).  The lines with
  circles and squares indicate where interference between the two
  trajectories in Fig.\ \ref{fig:geometry}b is constructive and
  destructive.}
\label{fig:grid}
\end{figure}

The interlayer ac conductivity $\sigma_c(\omega)$ is given by a
correlator of tunneling events at times $t$ and $t'$
\cite{McKenzie,Osada03}
\begin{equation}
  \sigma_c(\omega)\propto 
  {\mathcal Re}\, t_c^2 \left\langle \int_t^{\infty}
  e^{i\phi(t') - i\phi(t)}  e^{(t'-t)(i\omega-1/\tau)} 
  dt^{\prime} \right\rangle_t,
\label{tau}
\end{equation}
where $\tau$ is a relaxation time.  Substituting Eqs.\ (\ref{phi}) and
(\ref{xy}) in Eq.\ (\ref{tau}), we find
\begin{equation}
   {\sigma_c({\bm B},\omega)\over\sigma_c(0,0)} 
   = \sum_{n=-\infty}^\infty 
   \frac{J_n^2(B_x')}
   {1+(\omega_c\tau)^2(n-B_y'\mp\omega/\omega_c)^2},
\label{sigma_c}
\end{equation}
where $\sigma_c(0,0)$ is the dc conductivity at ${\bm B}=0$, and the
signs $\mp$ in the denominator originate from the $\pm v_F$ sheets of
the Fermi surface.  Eq.\ (\ref{sigma_c}) is in agreement with Refs.\ 
\cite{Yagi,Osada92,McKenzie,Osada03,Lebed03}.  It can be applied to
the microwave measurements at $\omega\neq0$
\cite{Ardavan98,Takahashi05}, but below we concentrate on the dc case
$\omega=0$.  When $\omega_c\tau\to\infty$, only the term with $n=B_y'$
survives, and Eq.\ (\ref{sigma_c}) reduces to $\sigma_c({\bm
  B},0)/\sigma_c(0,0)=(\tilde t_c/t_c)^2$ with $\tilde t_c$ from Eq.\ 
(\ref{J_n}).  In Fig.\ \ref{fig:grid}, we show the contour plot of
$\sigma_c({\bm B},0)/\sigma_c(0,0)$ vs.\ $B_x'$ and $B_y'$ calculated
from Eq.\ (\ref{sigma_c}) for $\omega_c\tau=\sqrt{50}\approx7.1$.  The
dc conductivity $\sigma_c$ is maximal at the vertical stripes, labeled
by the integer numbers $n$, which correspond to the Lebed magic angles
$B_y'=n$.  Within the $n$-th vertical stripe, $\sigma_c$ has
alternating maxima and minima, indicated by circles and squares, which
represent oscillations of $J_n^2$ vs.\ $B_x'$ in Eqs.\ (\ref{J_n}) and
(\ref{sigma_c}).  Positions of these maxima and minima can be obtained
from the Aharonov-Bohm interference in momentum space, as described
below.

Eqs.\ (\ref{H_perp}) and (\ref{phi}) show that, in the process of
interlayer tunneling, the in-plane electron momentum changes by
\begin{equation}
  {\bm q}=(q_x,q_y)=(ed/c)(B_y,-B_x).
\label{q}
\end{equation}
Thus, the Fermi surfaces of the two layers are displaced relative to
each other by the vector ${\bm q}$ \cite{McKenzie,Osada03}, as shown
in Fig.\ \ref{fig:geometry}b.  Electrons can tunnel between the layers
only at the intersection points ${\bm k}_1$, ${\bm k}_2$, ${\bm k}_3$,
etc.\ of the two Fermi surfaces, where the conservation laws of energy
and momentum are satisfied.  In the presence of $B_z$, there is a
phase difference between the two trajectories connecting the
intersection points, which is proportional to the shaded
momentum-space area $S_1>0$ or $S_2<0$ in Fig.\ \ref{fig:geometry}b.
The algebraic sum $S_1+S_2=q_x(2\pi\hbar/b)$ depends only on
$q_x=(ed/c)B_y$.  Constructive interference between ${\bm k}_1$ and
${\bm k}_3$ requires that $(S_1+S_2)c/\hbar eB_z=2\pi n$, which is
equivalent to the Lebed condition $B_y'=n$.

\begin{figure}
\includegraphics[width=0.8\linewidth]{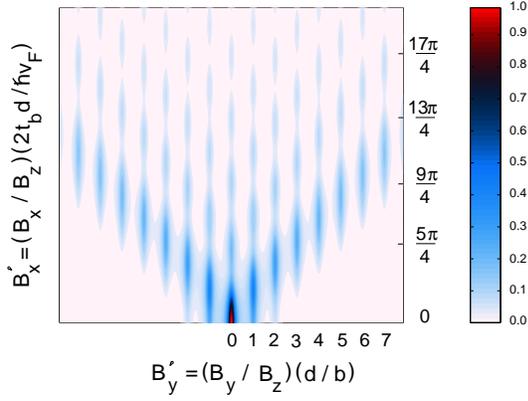}
\caption{Contour plot of the normalized interlayer dc conductivity 
  $\sigma_c({\bm B},0)/\sigma_c(0,0)$ vs.\ $B_x'$ and $B_y'$
  calculated from Eq.\ (\ref{sigma_c-m}).  Compared with Fig.\ 
  \ref{fig:grid}, this plot takes into account additional tunneling
  amplitudes $t_m$ along ${\bm d}+m{\bm b}$: $t_{\pm1}=t_c/2$ and
  $t_{\pm2}=t_c/4$, which produce the Lebed oscillations at $B_x=0$.}
\label{fig:sigma_m}
\end{figure}

Interference between ${\bm k}_1$ and ${\bm k}_2$ is controlled by the
area $S_1$.  Introducing the dimensionless variable $S_1'=S_1 c/\hbar
eB_z$, we find from Fig.\ \ref{fig:geometry}b that
\begin{equation}
  S_1'  = 
  2 B_x' \sqrt{1-\left(\frac{B_y'}{B_x'}\right)^2} +
  B_y'\left[\pi+2\arcsin\left(\frac{B_y'}{B_x'}\right)\right].
\label{S1}
\end{equation}
Constructive interference requires that $S_1'=2\pi(j+1/4)$, where $j$
is an integer, and the extra phase $\pi/2$ appears because ${\bm k}_1$
and ${\bm k}_2$ are the turning points on the Fermi surface, when
viewed along the vector $\bm q$.  The lines with circles show where in
Fig.\ \ref{fig:grid} this condition is satisfied.  Maxima of
$\sigma_c$ are achieved at the circled intersections of these lines
and the integer vertical lines, where both $S_1$ and $S_1+S_2$ give
constructive interference.  These points correspond to the maxima of
the Bessel functions in Eq.\ (\ref{sigma_c}).  The lines with squares
in Fig.\ \ref{fig:grid} show where the interference in $S_1$ is
destructive ($j$ is half-integer).  At the intersections of these
lines and the integer vertical lines, marked by squares, $\sigma_c$
has minima, and the Bessel functions in Eq.\ (\ref{sigma_c}) have
zeros.  There, $\sigma_c\to0$ at $\omega_c\tau\to\infty$, and
resistivity $\rho_c=1/\sigma_c$ increases without saturation when
$B\to\infty$, whereas $\rho_c(B)$ saturates at the circles
\cite{Lebed05}.  The maxima and minima of $\sigma_c$ create a
checkerboard pattern of oscillations \cite{Lebed03} for
$|B_x'|>|B_y'|$ in Fig.\ \ref{fig:grid}.

The diagonal line $B_x'=B_y'$ in Fig.\ \ref{fig:grid} corresponds to
the third angular effect \cite{Yoshino95,Osada96,Lebed97b}.  At this
line, the points ${\bm k}_2$ and ${\bm k}_3$ merge, and the area $S_2$
shrinks to zero in Fig.\ \ref{fig:geometry}b.  For $|B_x'|<|B_y'|$,
the two Fermi surfaces do not cross in Fig.\ \ref{fig:geometry}b, so
interlayer tunneling is suppressed, and $\sigma_c$ does not show
oscillations below the diagonal lines in Fig.\ \ref{fig:grid}.
However, this contradicts experiments \cite{Chaikin98a,Chaikin98b},
which show the Lebed oscillations of $\sigma_c$ vs.\ $B_y'$ at
$B_x=0$.

\begin{figure}
  \includegraphics[width=0.69\linewidth]{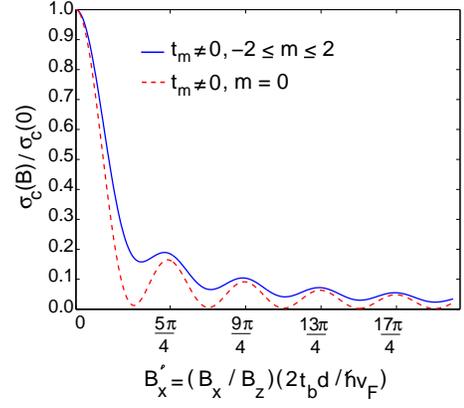}
\caption{Comparison between the normalized interlayer dc 
  conductivities $\sigma_c({\bm B,0})/\sigma_c(0,o)$ calculated from
  Eq.\ (\ref{sigma_c}) (dashed line) and Eq.\ (\ref{sigma_c-m}) (solid
  line) and plotted vs.\ $B_x'$ for $B_y=0$.  The DKC oscillations are
  reduced in the latter case, because of the additional tunneling
  amplitudes $t_m$.}
\label{fig:1Dsigma}
\end{figure}

To improve the theory, let us consider a model with interlayer
tunneling amplitudes $t_m$ between the chains shifted by $m$ units in
the $y$ direction \cite{Osada92}.  The tunneling
displacement is ${\bm d}+m{\bm b}$, so the phase in Eq.\ 
(\ref{H_perp}) becomes
\begin{equation}
  \phi({\bm r})=\frac{e}{\hbar c}(A_zd+A_ymb), \quad A_y=B_zx.
\label{phi_m}
\end{equation}
Comparing Eqs.\ (\ref{phi}) and (\ref{phi_m}), we see that results in
this case can be obtained by substitution $B_yd\to B_yd-B_zmb$ and
$B_y'\to B_y'-m$ in the old results. Eq.\ (\ref{J_n}) transforms into
$\tilde t_m=t_m\,J_{n-m}(B_x')$ for $B_y'=n$, and Eq.\ (\ref{sigma_c})
becomes
\begin{equation}
   {\sigma_c({\bm B},\omega)\over\sigma_c(0,0)} = 
   \sum_m \sum_{n=-\infty}^\infty 
   \frac{t_m^{\prime2} J_{n-m}^2(B_x')}
   {1+(\omega_c\tau)^2(n-B_y'\mp\omega/\omega_c)^2},
\label{sigma_c-m}
\end{equation}
where $t_m^{\prime2}=t_m^2/\sum_l t_l^2$.  The contour plot of Eq.\ 
(\ref{sigma_c-m}) can be obtained by shifting the plot in Fig.\ 
\ref{fig:grid} by $m$ units along the $B_y'$ axis and adding the
shifted plots with the weights $t_m^{\prime2}$.  The resulting contour
plot, calculated for $t_0=t_c$, $t_{\pm1}=t_c/2$, and
$t_{\pm2}=t_c/4$, is shown in Fig.\ \ref{fig:sigma_m}.  At $B_x=0$,
$\sigma_c(B_y')$ has maxima for those directions $B_y'=m$ where $t_m$
exists \cite{Osada92,Chaikin02}.  Oscillations of $\sigma_c$ vs.\ 
$B_x'$ are smeared in Fig.\ \ref{fig:sigma_m}, because the shifted
maxima and minima of the checkerboard pattern in Fig.\ \ref{fig:grid}
add up out of phase.  This is illustrated in Fig.\ \ref{fig:1Dsigma},
which shows that the DKC oscillations of $\sigma_c(B_x')$ for $B_y=0$
are much weaker for multiple $t_m$.  Moreover, $\sigma_c$ does not
have zeros at $B_y'=n$.  Thus, when $B\to\infty$, $\rho_c(B)$
saturates on the integer lines in Fig.\ \ref{fig:sigma_m}, but grows
without saturation between the lines.  Weak DKC oscillations of
$\sigma_c(B_x')$ at $B_y=0$ and strong Lebed oscillations of
$\sigma_c(B_y')$ at $B_x=0$ correspond qualitatively to
(TMTSF)$_2$PF$_6$ \cite{Chaikin95c,Chaikin98a,Chaikin98b}, indicating
that several $t_m$ are present.  The opposite case, strong DKC and
weak Lebed oscillations, is found in $\rm(TMTSF)_2ClO_4$
\cite{Osada91,Naughton91,Chaikin94a}, suggesting that it has only one
dominant $t_0=t_c$ \cite{Takahashi05}.  The model parameters can be
determined by quantitative comparison between the calculated plots and
experimental data for $\sigma_c(B_x',B_y')$.  Fig.\ \ref{fig:sigma_m}
shows that the strength of the Lebed oscillations in $\sigma_c$ vs.\ 
$B_y'$ increases when $B_x'\neq0$, in agreement with the Lee-Naughton
experiment \cite{Naughton98a}.

The amplitudes $t_m$ do not necessarily represent electron overlap
between distant chains.  They may be effective parameters in a model
\cite{Lebed04}, where $\varepsilon(k_x)$ has curvature, so $v_F$
depends on $k_x$ and varies along the quasiclassical trajectory
(\ref{xy}).  The resulting expression for $\sigma_c$ has a form
similar to Eq.\ (\ref{sigma_c-m}) with some effective parameters
$t_m$, which themselves may depend on $\bm B$ \cite{Lebed04}.

While Eq.\ (\ref{sigma_c-m}) may well describe the oscillatory part of
$\sigma_c$, it often fails to describe the background, particularly in
$\rm(TMTSF)_2PF_6$ \cite{Chaikin95c,Chaikin98a,Chaikin98b,Chaikin02},
although there are variations with pressure and sample
\cite{Naughton98b}.  This remains one of the open problems, along with
unusual temperature dependence of resistivity \cite{Chaikin98a} and
mysterious angular oscillations of the Nernst effect \cite{Nernst}.

We presented a unified geometrical explanation of different types of
AMRO in Q1D conductors in terms of Aharonov-Bohm interference in
interlayer electron tunneling.  We visualized a two-parameter pattern
of oscillations for generic magnetic field orientations using the
natural variables $B_x'$ and $B_y'$.  Quantitative comparison with
experimental data plotted in this way is needed.  This work was
supported by the NSF Grant DMR-0137726.

\vspace{-\baselineskip}

\end{document}